\documentclass[12pt]{article}

\usepackage{times}

\usepackage{graphicx}

\newenvironment{sciabstract}{%
\begin{quote} \bf}
{\end{quote}}

\newcounter{lastnote}

\title{On-Chip Microwave Quantum Hall Circulator} 

\author
{A. C. Mahoney,$^{1,2\ast}$ J. I. Colless,$^{1,2\ast}$ S. J. Pauka,$^{1,2}$ J. M. Hornibrook$^{1,2}$,\\
J. D. Watson$^{3,4}$, G. C. Gardner$^{4,5}$, M. J. Manfra$^{3,4,5}$, A. C. Doherty$^{1}$, and D. J. Reilly$^{1,2\dagger}$\\
\\
\normalsize{$^{1}$ARC Centre of Excellence for Engineered Quantum Systems,}\\
\normalsize{School of Physics, The University of Sydney, Sydney, New South Wales 2006, Australia.}\\
\normalsize{$^{2}$ Microsoft Station Q Sydney, The University of Sydney, Sydney, New South Wales 2006, Australia.}\\
\normalsize{$^{3}$Department of Physics and Astronomy, Purdue University, West Lafayette, Indiana 47907, USA.}\\
\normalsize{$^{4}$Birck Nanotechnology Center, School of Materials Engineering and School of Electrical}\\
\normalsize{and Computer Engineering, Purdue University, West Lafayette, Indiana 47907, USA.}\\
\normalsize{$^{5}$Microsoft Station Q Purdue, Purdue University, West Lafayette, Indiana 47907, USA.}\\
\\
\normalsize{$*$These authors contributed equally to this work.}
\\
\normalsize{$^\dagger$To whom correspondence should be addressed; E-mail:  david.reilly@sydney.edu.au}
}
\date{}

\begin{document} 

\baselineskip20pt

\maketitle 

\begin{sciabstract}
Circulators are non-reciprocal circuit elements integral to technologies including radar systems, microwave communication transceivers, and the readout of quantum information devices. Their non-reciprocity arises from the interference of microwaves over the centimetre-scale of the signal wavelength in the presence of bulky magnetic media that break time-reversal symmetry.
 Here we realize a completely passive on-chip microwave circulator with size 1/1000th the wavelength by exploiting the chiral, `slow-light'  response of a 2-dimensional electron gas (2DEG) in the quantum Hall regime. For an integrated GaAs device with 330 $\mu$m diameter and $\sim$ 1 GHz centre frequency, a non-reciprocity of 25 dB is observed over a 50 MHz bandwidth. Furthermore, the direction of circulation can be selected dynamically by varying the magnetic field, an aspect that may enable reconfigurable passive routing of microwave signals on-chip. 
\end{sciabstract}

\clearpage
Miniaturized, non-reciprocal devices are currently of broad interest for enabling new applications in acoustics \cite{Fleury31012014}, photonics \cite{Feng05082011,Bi_2011}, transceiver technology \cite{Alu_Nat_Phys}, and in the regime of near quantum-limited measurement \cite{PhysRevLett.93.126804,PhysRevApplied.4.034002,PhysRevX.3.031001,PhysRevX.5.041020,DiVincenzo}, where they are needed to isolate qubits from their noisy readout circuits.  
Since the 1950s, passive circuit elements exhibiting non-reciprocity at microwave frequencies have been implemented using bulky magnetic devices that are comparable in scale to the centimetre wavelength of signals in their operating band. The footprint of these components poses a major limitation to integrating entire systems on a chip, such as what is required, for instance, to scale-up quantum computing technology. 

A seemingly obvious means of realizing non-reciprocity on a semiconductor chip is to use the Hall effect, where an external magnetic field breaks the time reversal symmetry of electrical transport \cite{Mason}. Hall-based devices were ruled out in 1954 however \cite{Wick}, since near the electrical contacts, where the current and voltage are collinear, dissipation is so significant that the usefulness of this approach is greatly limited. This dissipative mechanism has an analog in the quantum Hall regime where the two-terminal resistance of a device is always finite over a scale of the inelastic scattering length as carriers transition from their contacts to the dissipationless, one-dimensional (1D) edge-states that support transport \cite{PhysRevB.38.9375}. Recently, Viola and DiVincenzo \cite{DiVincenzo} have proposed a means of addressing the limitation imposed by 2-terminal dissipation, suggesting the possibility of coupling microwave signals to the edge of a quantum Hall device reactively, without using traditional ohmic contacts. In a geometry where the signal ports of the device are positioned orthogonal to an incompressible quantum Hall edge-state, microwave power is coupled capacitively and non-dissipative transport in one-direction appears possible \cite{DiVincenzo}. 

Here we engineer, on-chip, a chiral microwave interferometer that yields a high degree of non-reciprocity and dynamic range, with the low-dissipation inherent to edge transport in the quantum Hall regime. Configured as a completely passive 3-port circulator, our device demonstrates non-reciprocal operation at frequencies and magnetic fields commonly used for the read out of spin qubits \cite{Reilly:2007ig,Barthel:2009hx,Colless_PRL}, facilitating integration with such semiconductor technologies. In comparison to traditional ferrite-based microwave components, this quantum Hall circulator is reduced in size by a factor $\sim$ 1/1000th the wavelength and exhibits a new mode of operation in which the direction of circulation can be dynamically reconfigured by altering the strength of the magnetic field. A simple model based on a Fano-resonance mechanism \cite{RevModPhys.82.2257} qualitatively accounts for the observed phenomena. 

Central to the operation of our device are edge magnetoplasmons (EMPs) \cite{volkov1988edge} that propagate along a quantum Hall edge in response to a capacitively coupled microwave excitation \cite{TALYANSKII199040, ashoori1992edge,Zhitenev,PhysRevLett.113.266601}. These chiral excitations travel with a velocity $v_{EMP}\sim|\vec{E}|/|\vec{B}|$,  set by the ratio of the electric field $\vec{E}$ at the sample boundary and the applied magnetic field $\vec{B}$ \cite{TALYANSKII199040}. For a high mobility 2DEG formed at the interface of the semiconductors GaAs and AlGaAs (see supplementary materials for details), the velocity of the EMP modes is typically $v_{EMP}\sim $10$^{5}$ ms$^{-1}$ \cite{PhysRevB.84.045314, PhysRevB.81.085329}, some 1000 times slower than the speed of light in the semiconductor dielectric. In order to exploit these EMPs to realize non-reciprocal microwave devices, we first detect their presence in a contactless etched disk of quantum Hall fluid by coupling to a proximal metallic coplanar transmission line (CTL) \cite{PhysRevB.88.165305}, as shown in Fig. 1A and 1B. By measuring the transmitted microwave power through the transmission line as a function of frequency $f$, a spectrum of discrete features is observed with applied magnetic field $B$ (Fig. 1C). We identify EMP modes in the data with frequencies set by the edge velocity and circumference of the disk, following the dependence $f\sim B^{-1}$(log($B^2$) + const.) \cite{volkov1988edge}, consistent with the dielectric constant of GaAs \cite{ashoori1992edge,Balaban} (see supplementary materials). Comparing the microwave spectrum to transport measurements from a Hall-bar on the same chip (Fig. 1D), we note that at high field (with only the last few Landau levels occupied) features resolve into discrete, crescent-shaped resonances that coincide with minima in the longitudinal resistance $R_{xx}$, where dissipation is suppressed.
\clearpage
\noindent\includegraphics[trim = 0mm 0mm 0mm 0mm, width=15cm]{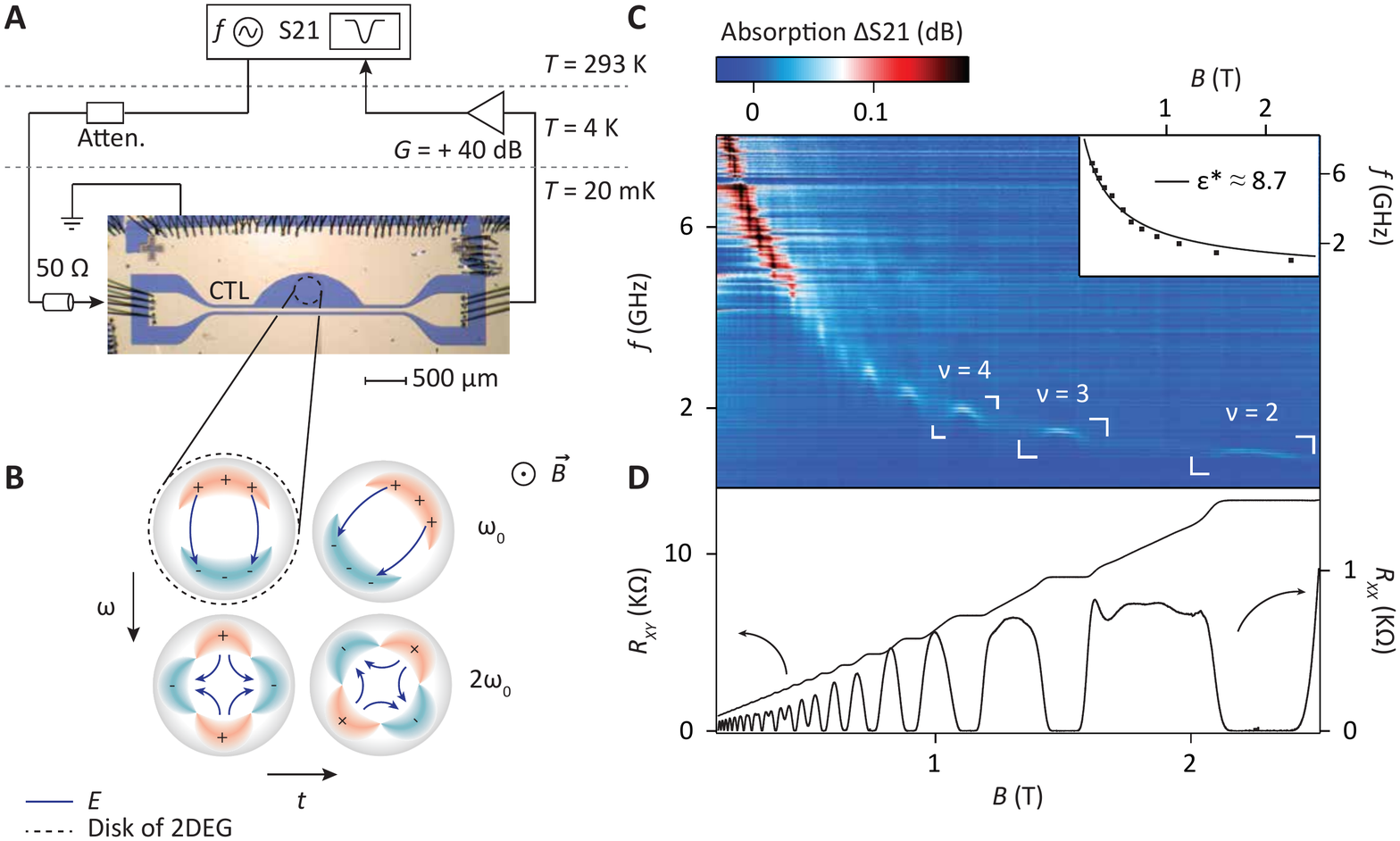}\\
\noindent {\bf Figure 1: Detecting microwave edge magnetoplasmons (EMPs).} 
{\bf{(A)} } Experimental setup including photograph of a coplanar transmission line device similar to that used to perform measurements coupled to a 350 $\mu$m etched disc of 2DEG (black dashed circle) at fridge temperature $T$ = 20 mK. A vector network analyser is used to excite EMP modes across a wide frequency range and microwave absorption is measured as the ratio of the amplified output to input signal ($S_{21}$) from the CTL. 
{\bf{(B)} }
Illustration of the fundamental (top row) and first harmonic (bottom row) EMP modes as they evolve with time, where $\omega_0$ is the fundamental mode and 2$\omega_0$ the first harmonic (adapted from \cite{volkov1988edge}). Charge distributions and electric fields $\vec{E}$ are indicated schematically. An external magnetic field $B$ applied to the device points out of the page.
{\bf{(C)} }
EMP spectrum of the quantum Hall disk showing absorbed microwave power as a function of frequency and magnetic field. Data has had a background, obtained at high field, subtracted. Inset shows the position of absorption dips at integer quantum Hall filling factors. Black line is a fit that allows an average dielectric constant of $\epsilon^{*} \approx 8.7$ to be extracted, consistent with excitations of an edge-state in GaAs (see supplementary material). 
{\bf{(D)}} 
Transverse ($R_{xy}$) and longitudinal ($R_{xx}$) Hall resistance measurements taken at $T$ = 20 mK on a Hall bar proximal to the microwave disk. The 2DEG is 270 nm below the surface with carrier density n$_s$ = 1.1 x 10$^{11}$ cm$^{-2}$, and mobility $\mu$ = 5.2 x 10$^6$ cm$^2$/Vs. 
\newline

To test if these edge magnetoplasmons support the non-reciprocal transmission of microwaves, we implement a standard circulator configuration, with 3 ports arranged at 120-degree intervals around a disk of 2DEG (330 $\mu$m diameter), as shown in Fig. 2A and 2B. For a single edge at high magnetic field, a voltage applied to a port capacitance induces an orthogonal current in the edge-state, with an impedance of the order of the inverse conductance quantum ($\sim$ 26 k$\Omega$). Given our measurement setup uses electronic components with a characteristic impedance of $Z_0 \sim$ 50 $\Omega$, we have added an impedance matching circuit to enhance the response of each port (our matching network is a series chip-inductor $L$ = 47 nH in resonance with the stray capacitance $C_{\rm stray}$, see supplementary materials). The circulator is further embedded in a reflectometry configuration (see Fig 2C) that enables a measurement of the port reflection as well as port transmission coefficient, from which dissipation can be estimated. As a control we first measure all microwave $S$-parameters at zero magnetic field, observing that all directions and ports are equivalent, as shown in Fig. 2D.  An overall frequency-dependent, but reciprocal response can be associated with the impedance matching network, with matching frequency set to $1/\sqrt{LC_{\rm stray}} \sim$ 1 GHz.  All subsequent measurements are normalized relative to this zero-field transmission response.
\clearpage
\noindent
\noindent\includegraphics[trim = 0mm 0mm 0mm 0mm, width=\textwidth]{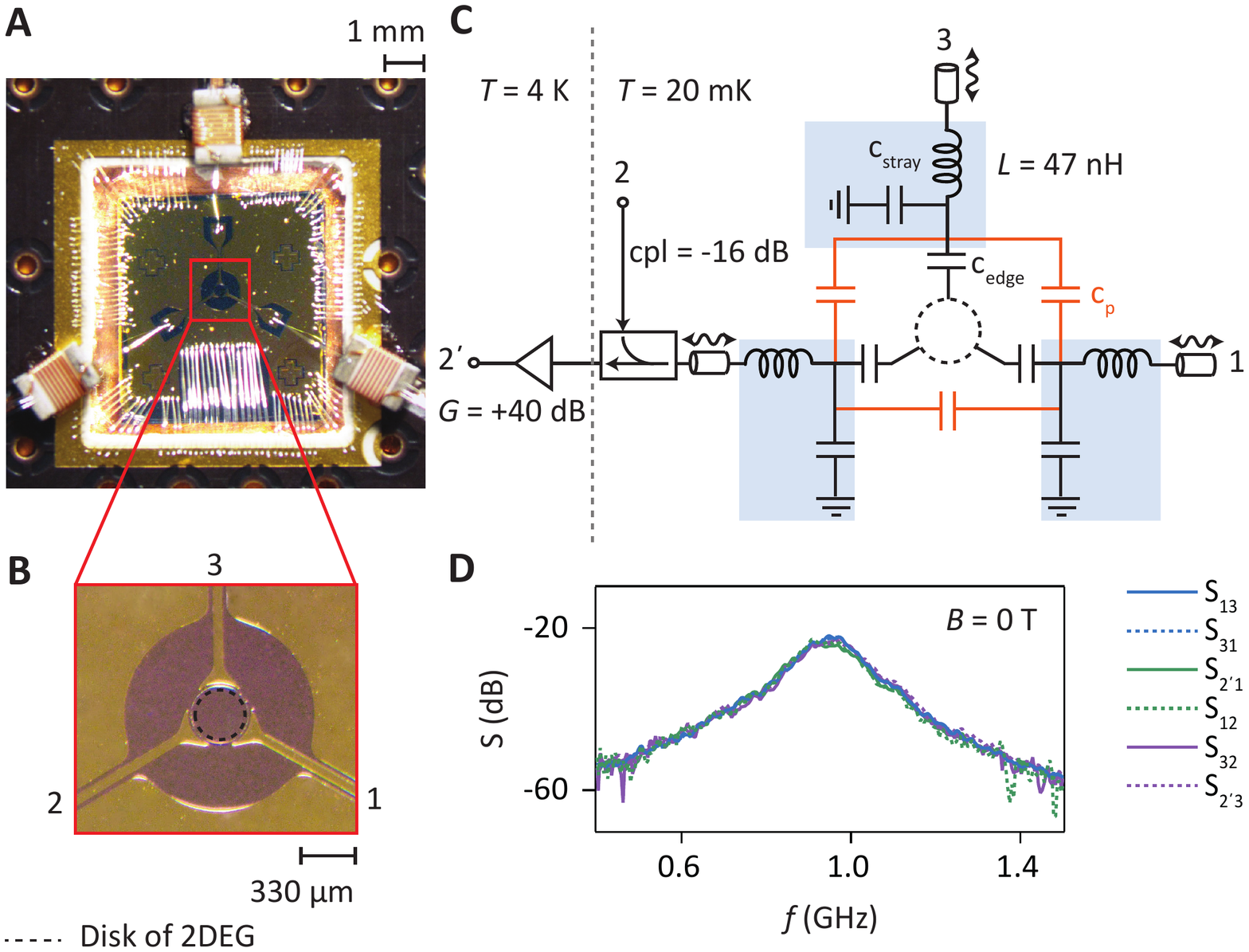}
\noindent {\bf Figure 2: Experimental setup for determining the response of the on-chip circulator.} {\bf{(A)}}  Photograph of circulator device showing the three coplanar transmission lines connected to copper chip-inductors (47 nH) for impedance matching. {\bf{(B)}} Close-up of false-coloured photo of the circulator showing 330 $\mu$m diameter 2DEG disc with a 20 $\mu$m gap to the metal defining the three signal ports. {\bf{(C)}} Circuit schematic of the experimental setup indicating port-to-edge capacitive coupling $C_{\rm edge}$ and direct parasitic coupling between ports $C_{\rm p}$. Resonant ($LC_{\rm stray}$) matching circuits are indicated with blue boxes. The input of port 2 passes through a directional coupler, with the reflected signal coupled to the output line (denoted 2$^\prime$) and amplified. {\bf{(D)}} Shows the full 6-way transmission response of the circulator at zero magnetic field, with $S$-parameter measurements indicating complete reciprocity and a frequency response that arises from the matching networks. For each port the measured response of the amplifiers, couplers and cold attenuators in the circuit have been subtracted.
\newline

Turning to our key result, Figure 3 shows the full transmission response of the 3-port circulator in the presence of a magnetic field that breaks time-reversal symmetry. Similar to the EMP spectrum of Fig. 1C, we first observe the presence of EMPs that enhance the transmitted power at certain frequencies, broadly following an approximate $f \sim B^{-1}$ dispersion relation, as is seen in Fig. 3A ($S_{13}$) and 3B ($S_{31}$). Strikingly, there are regions of the spectrum where the transmitted power appears to flow in either a forward or reverse direction with respect to the chirality of the edge. Particularly apparent are the crescent-shaped features that switch from forward to reverse transmission at distinct frequencies. This phenomenon, with a peak near the fundamental frequency of the EMP mode and a dip near the first EMP harmonic, is seen for all $S$-parameters in the chiral (clockwise) direction of the 3-port device (see solid lines in Fig. 3D). 
\newline
To measure the extent of non-reciprocity in our circulator, Figure 3C shows the difference between forward and reverse power by subtracting $S_{31}$ from $S_{13}$. Unlike the $B$ = 0 data shown in Fig. 2D, we now observe a strong directional dependence in the isolation between ports, that approach 40 dB at particular frequencies and magnetic fields (Fig. 3F). Alternatively, we can also test for non-reciprocity by comparing the response of signals from two different inputs of the circulator to a common output. Since the device is geometrically symmetric, the response from the separate paths $S_{2'1}$ and $S_{2'3}$ are the same at $B$ = 0, (see Fig. 2D). In the presence of a magnetic field however, Fig. 3G shows that these paths are no longer equivalent, but depend rather on the direction of the field. This is clear in the data (Fig. 3G), since blue and red features are not mirrored about $B$ = 0. 
\clearpage
\noindent
\noindent\includegraphics[trim = 0mm 0mm 0mm 40mm, width=\textwidth]{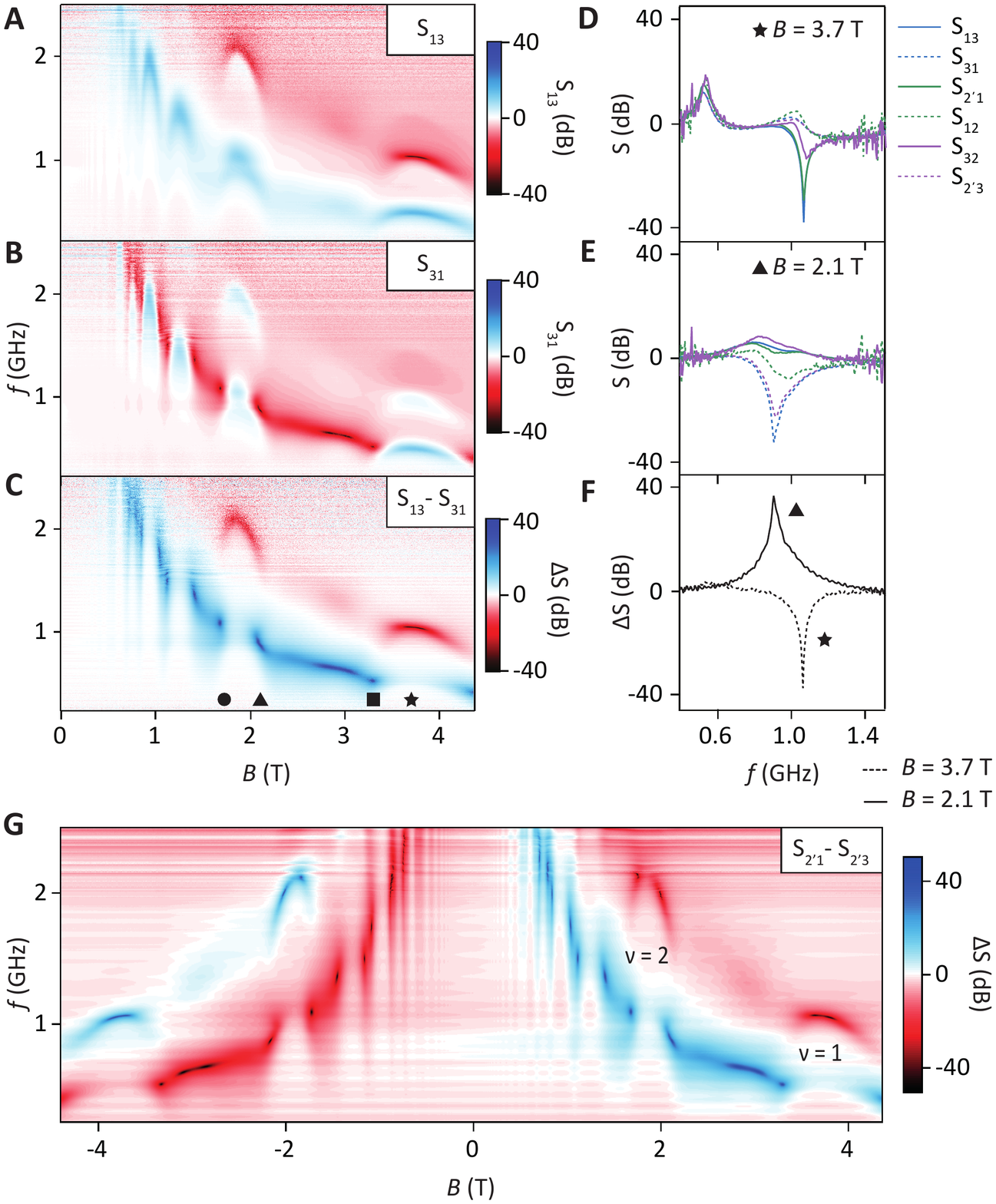}
\noindent {{\bf Figure 3: The non-reciprocal response of the quantum Hall circulator.}} {{\bf{(A and B)}}} Port transmission $S_{13}$ and $S_{31}$ with frequency and magnetic field. All measurements have been normalized to the gain-corrected background at $B$ = 0 (shown in Fig. 2{\bf{D}}), which defines the 0 dB colour scale. {\bf{(C)}} Differential microwave response $S_{13}-S_{31}$ showing strong, frequency and $B$-dependent non-reciprocity. {\bf{(D and E)}} Show the full combination of transmission $S$-parameters, taken at $B$-fields indicated by the symbols in {\bf{C}}. {\bf{(F)}} shows slices through the colour scale data in {\bf{C}} demonstrating forward and reverse circulation. {\bf{(G)}} Isolation, $S_{2'1}$-$S_{2'3}$ measured at positive and negative magnetic fields. Note the anti-symmetry of the features with respect to the $B$ = 0 axis.
\newline

Comparing the microwave response of the circulator to independent quantum Hall transport data suggests  two distinct regimes. Between integer filling factors, where $R_{xx}$ is maximised in transport, there is a large non-reciprocity in the microwave response, but also likely strong dissipation. Contrasting these broad regions are narrow crescent-shaped features that occur at fields corresponding to integer filling. These narrow features are particularly strong at frequencies near twice the fundamental EMP resonance. Again, overlaying these features with transport measurements on the Hall bar indicates they align with minima in $R_{xx}$, where dissipation is suppressed. Microwave loss measurements, taken by summing the transmitted and reflected signal in comparison to the incident power, are consistent with a level of dissipation set by $R_{xx}$ as measured in transport. 
\clearpage
\noindent
\noindent\includegraphics[width= \textwidth]{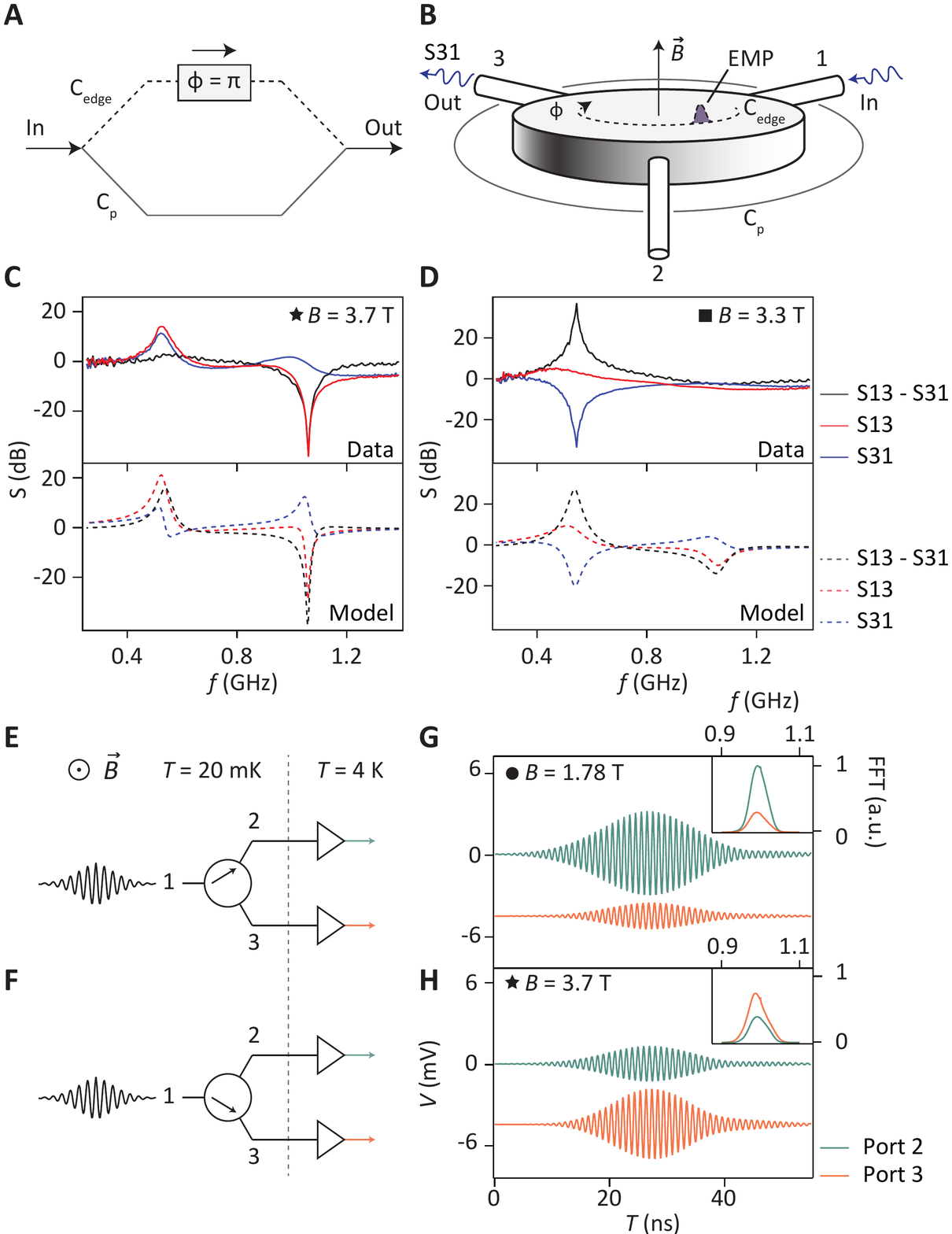}
\newline
\noindent {{\bf Figure 4: Forward and reverse circulation.}} {{\bf{(A)} }and {\bf{(B)}} Shows an  interferometric mechanism underlying operation of the quantum Hall circulator, with slow plasmonic path, via $C_{\rm edge}$, and the direct capacitive path, via $C_{\rm p}$. Non-reciprocal isolation between ports is produced for frequencies where the two paths are out of phase by $\phi = \pi$ in the forward direction and $\phi = 2\pi$ in the reverse direction. {{\bf{(C)} }and {\bf{(D)}} compare a simple model that captures this physics to the experimental data at magnetic fields indicated by the symbols (with respect to Fig. 3C). The model is described in detail in the supplemental material with parameters set to values $Z_0$ = 50 $\Omega$, $C_{\rm p} = 315$ fF, $C_{\rm edge} = 127$ fF, $R_{xy}$ = 5000  $\Omega$ with $R$ = 30 $\Omega$ in the centre of an EMP resonance (star symbol) and $R$ = 350 $\Omega$ off resonance (square symbol). {{\bf{(E)} }and {\bf{(F)}} Schematic showing the experimental configuration for reconfigurable routing of microwave signals on-chip. A wavepacket is directed to port 1 of the circulator, and the resultant signals are measured after amplification at ports 2 and 3. {{\bf{(G)} }and {\bf{(H)}} shows the output signal response, as configured by adjusting the amplitude of the magnetic field to direct packets to the required port. Insets show normalized fast Fourier transform (FFT) amplitudes. 
\newline

We account for the distinct features in our measurements, as well as the phenomena of forward and reverse circulation via a simple picture of a Fano-like resonance. Figure 4 illustrates the phenomenology of the quantum Hall circulator. Similar to the operation of a traditional ferrite device, we consider a resonator structure with two interfering paths, as shown in Fig. 4A. The arms of this interferometer comprise a direct path, supported by the parasitic (geometric) capacitance $C_{\rm p}$ between ports, and an indirect path $C_{\rm edge}$, that capacitively couples ports via the plasmonic excitation of a quantum Hall edge. Key to the operation of our circulator is this `slow light' response of the EMP modes, which, traveling at velocities 1000-times slower than the microwaves in the direct path, acquire the same phase over a length scale that is 1000-times shorter than the microwave wavelength in the dielectric. Considering these two-paths we note that there will be a frequency near the EMP resonance, at which the phase acquired via the edge leads to complete destructive interference with the signal propagating via the direct path. Given the chirality of the EMP, the condition for destructive interference will be dependent on the direction of microwave transmission, producing a non-reciprocal response between adjacent ports. Take for instance, the case where signals from port 3 to 1 propagate clockwise via the edge capacitance $C_{\rm edge}$ and acquire a phase of $\pi$-radians with respect to the signal traveling via $C_{\rm p}$. Interference of these signals isolates port 1, whereas reverse transport, from port 1 back to port 3 must continue in a clockwise direction, past port 2 and acquire a constructive phase of $2 \pi$ over twice the length. Circulation in the opposite direction to the chirality of the edge can now be understood for frequencies in which a $\pi$-phase is acquired in the forward direction, but $2 \pi$-phase in reverse.

We construct a simple model based on this Fano-like picture of interfering paths \cite{RevModPhys.82.2257}, by modifying the standard response of a three-terminal Carlin circulator to account for transport via a quantum Hall edge (see supplemental material). This yields an expression for the non-reciprocal admittance matrix of the edge, $Y_{\rm edge}$, as was done in Ref. \cite{DiVincenzo}. Extending the model in Ref. \cite{DiVincenzo}, we add an additional admittance term $Y_{\rm p}$ to account for a direct parasitic coupling $C_{\rm p}$ between terminals (see Fig. 2B). We further include the possibility of loss $R$, either directly in the chiral EMP mode or elsewhere in the circuit. Given an admittance of the edge-state $Y_{\rm edge}$, the total admittance is then given by: ${Y_{{\rm{total}}}} = {({I} + R  Y_{\rm edge})^{ - 1}}Y_{\rm edge} + {Y_{{\rm{p}}}}$, where $I$ is the identity matrix and where \[ {Y_{{\rm{p}}}} = \left( {\begin{array}{*{20}{c}}
{2c}&{ - c}&{ - c}\\
{ - c}&{2c}&{ - c}\\
{ - c}&{ - c}&{2c}
\end{array}} \right)\]\vspace{0.01 in}
with $c = i\omega {C_p}$ and $\omega$ is the angular frequency of the microwaves. Microwave $S$-parameters can then be calculated as a function of $\omega$ for a given characteristic impedance of the input port ($Z_0$).

This model qualitatively captures the mechanism of circulation as arising from the interference of the parasitic and quantum Hall edge paths. Despite its simplicity, we find it also accounts for many of the features seen in the experimental data, including forward and reverse circulation that occurs near the fundamental and first harmonic of the EMP mode, as shown in Fig. 4C and 4D. For features that occur at fields corresponding to integer filling, we find good agreement with the data for parameter values that are consistent with the device geometry and independent transport measurements (see Fig. 4 caption for details). At magnetic fields slightly away from integer filling, increasing $R$ in the model yields good agreement with the data.  

Finally, having established the mechanism leading to non-reciprocity in our device, we turn to describe a new mode that has no analog in the operation of classical circulators but may enable reconfigurable passive routing of microwave signal on-chip. To demonstrate this mode, we generate a microwave packet with frequency $\sim$ 1 GHz (see supplementary material), and feed it to port 1 of our circulator. Selective routing of this packet to either port 2 or port 3 is then controlled by selecting the magnetic field that produces forward or reverse circulation, as shown in Fig. 4E-H. We note that although, in this instance, the velocity of the EMP (and thus the acquired phase) is configured with an external magnetic field, similar selectivity is possible using a metallic surface gate to dynamically modify the edge length, carrier density, or electric field at the disk boundary \cite{KamataH:eo,Kumada:2013hk}. Considering the edge-state as a mesoscale delay-line with dynamic, gate-tunable length opens the prospect of compact, parametric devices such as amplifiers and mixers based on the plasmonic and chiral response of the quantum Hall effect. Indeed, such modes can also likely be realized at zero magnetic field using topological insulator devices that exhibit the quantum anomalous Hall effect \cite{Chang12042013}. 
\newline

We thank David DiVincenzo for useful conversations. This research was supported by Microsoft Research,  the US Army Research Office grant W911NF-14-1-0097, and the Australian Research Council Centre of Excellence Scheme (Grant No. EQuS CE110001013).
\bibliographystyle{science}

\end{document}